\documentclass[aps,twocolumn]{revtex4}
\usepackage{amsmath}
\usepackage{graphics}

\newcommand{\sidef}[1]{\left<#1\right>}
\newcommand{\sub}[1]{_{\text{#1}}}
\begin{document}

\title{Coil-Globule Transition for Regular, Random and Specially
Designed Copolymers: Monte Carlo Simulation and Self-Consistent Field Theory.}

\author{J. M. P. van den Oever, F. A. M. Leermakers and G. J. Fleer}
\affiliation{Laboratory of Physical Chemistry and Colloid Science, \\ Wageningen University, Dreijenplein 6,       \\ 6703 HB Wageningen, The Netherlands}

\author{V. A. Ivanov, N. P. Shusharina and A. R. Khokhlov}
\affiliation{Physics Department, Moscow State University, \\ Moscow 117234, Russia}

\author{P. G. Khalatur}
\affiliation{Department of Physical Chemistry, Tver State University,\\ Tver 170002, Russia}

\date{\today}

\begin{abstract}
The coil-globule transition has been studied for A-B copolymer chains both by means of lattice Monte Carlo simulations using bond fluctuation algorithm and by a numerical self-consistent field method. Copolymer chains of fixed length with A and B monomeric units with regular, random and specially designed (protein-like) primary sequences have been investigated. The dependence of the transition temperature on the AB sequence has been analyzed. A protein-like copolymer is more stable than a copolymer with statistically random sequence. The transition is more sharpe for random copolymers. It is found that there exist a temperature below which the chain appears to be in the lowest energy state (ground state). Both for random and protein-like sequences and for regular copolymers with a relatively long repeating block, a molten globule regime is found between the ground state temperature and the transition temperature. For regular block copolymers the transition temperature increases with block size. Qualitatively, the results from both methods are in agreement. Differences between the methods result from approximations in the SCF theory and equilibration problems in MC simulations. The two methods are thus complementary.
\end{abstract}

\maketitle

\section{Introduction}

Macromolecules composed of a single type of segment, i.e., homopolymers, behave as fractal objects: the radius of gyration $R\sub{g}$ of the molecule scales \cite{flory,edwa1965,book} with $N$ repeating units along the chain as $R\sub{g} \propto N^{\alpha}$, where $\alpha< 1$. The scaling exponent depends on the polymer concentration and the solvent quality. When the chains are in the dilute regime, i.e. when the inter-chain distance exceeds the coil size, the exponent decreases with decreasing solvent quality. In a good solvent the chains are highly swollen, $\alpha\approx0.6$. Under ideal or $\theta$-conditions the coils are Gaussian and $\alpha= 0.5$, whereas upon worsening the solvent quality below the $\theta$-point the coil collapses to a globule with $\alpha= 1/3$. This last transition is referred to as the coil-globule transition. For homopolymers the collapse transition is directly followed by macroscopic phase separation because the collapsed chains also attract each other.

For macromolecules composed of more than one segment type, i.e., copolymers or heteropolymers, the coil-globule transition and phase separation are usually not directly coupled. The uncoupling occurs especially when the solvent is selective for the various units along the chain. The segments for which the solvent is poor will cluster into a dense core and solvated segments will tend to accumulate on the outside of the globule. These segments on the outside may protect the system from macroscopic phase separation. In principle, some inter-chain aggregation remains possible. Similarly as in, e.g., surfactant systems, copolymer chains may aggregate into aggregates of finite size. If the protection process is sufficiently effective, inter-chain aggregation may be completely stopped and uni-molecular objects are formed which have a dense core and a solvated corona. Typical biological examples are globular proteins.

The coil-globule transition in heteropolymers has been studied by many authors in recent years using both analytical techniques and computer simulations
\cite{paper1,paper3,paper5,tkd1998,dbcm1999,krdkt1998,gana1998,nzo1998c}.
One of the motivations for these studies is the direct link to protein stability. Questions regarding the conformational stability of these molecules in bulk and the corresponding (in)stability at liquid-liquid or at solid-liquid interfaces remain largely unanswered as yet. One of the reasons for this is that there are few theoretical approaches that can be used to systematically and effectively investigate these phenomena.

The purpose is to launch a line of research aimed at resolving some of the issues regarding the conformational properties of heteropolymers, especially in interfacial systems. In the first step it is necessary to select theoretical tools and appropriate models for these purposes. The testing ground for such an investigation is the study of the coil-globule transition of heteropolymers in a dilute solution. The next step is to apply similar tools to interfacial systems. For obvious reasons it is not workable to include the full complexity of real block copolymer systems. On the other hand it is necessary to include in the model the fundamental features. Arguably, the essential first step is to use heteropolymers with two type of polymer units, hydrophilic (A) and hydrophobic (B), as model molecules. Therefore, the coil-globule transition in dilute bulk solution of three types of heteropolymers: regular copolymers, random copolymers and `protein-like' \cite{paper1,paper5} copolymers was investigated. Specifically we are interested in the dependencies of the transition temperature on the primary AB sequence distribution. In particular we are aimed to find the influence of the primary structure on critical properties of copolymer chains and the role of long-range correlations in the primary structure of protein-like copolymers. Using two computation techniques for investigating the problem of the coil-globule transition for copolymers with pre-set primary sequence, we can discover how both methods perform for this problem and how they complement each other.

The two models used are necessarily rather primitive. Efficiency considerations prompt us to use models that employ lattice approximations. More specifically we use a numerical self-consistent field (SCF) theory and Monte Carlo (MC) simulations. The motivation for the use of SCF tools is that the most sophisticated models of polymer adsorption are of this type\cite{sf,fleerbook}. The SCF method is based upon an approximate partition function of the system from which all corresponding thermodynamic quantities are computed directly. Not all excluded-volume correlations are included: the polymer molecule is modeled as a Markov chain. Especially for the study of proteins that have a unique tertiary structure, SCF techniques are not optimal. However, we will show that the current SCF technology can be extended so that some important features of the conformations of heteropolymers are within reach.

The motivation for using Monte Carlo simulations, and especially the bond-fluctuation model \cite{bfm1a,bfm1b,bfm2a,bfm2b}, is that this method has successfully been used to analyze various aspects of the coil-globule transition in homopolymer \cite{wkg1996,ny1998,ipb1998,isvpb2000} as well as heteropolymer \cite{paper1,paper3,paper5} chains. By means of this model the adsorption of copolymer chains with different primary structures on a flat surface has been also studied \cite{zkk1999}. A strong point of MC simulations is that all the excluded volume correlations are included: the chain is fully self-avoiding. This type of exactness has its price: a statistically sound solution is achieved only after a very large number of trial steps. In other words, the simulations are very time consuming and in practice it is not always possible to equilibrate the system in all aspects; the sampling of conformations in densely packed systems is difficult. We have used a multiple histogram reweighting technique \cite{histrw0,histrw0a,histrw} to extrapolate to the behavior of the chains away from the transition region \cite{ny1998,ipb1998,isvpb2000,histrw2}. Well equilibrated points near the coil to globule transition were used for this. This approach proved to give satisfactory results. More efficient algorithms for sampling the conformational statistics in dense systems will be needed to obtain direct results for the dense globules. Such improvements have been suggested recently \cite{pmb}. In this paper we have used a standard implementation of the bond fluctuation model. Some thermodynamic quantities are readily computed, such as the average energy in the system, or the mean size of the polymer chain. Other quantities, such as the entropy and the free energy, are more expensive. We  therefore have concentrated on the former properties in comparing the models.

The remainder of this paper is as follows. In the next section we will specify the necessary details of the two methods. Then we present some key results for both methods. In the discussion the focus will be two-fold. One of the goals is to obtain more knowledge of the coil-globule transitions for various types of copolymers. The other goal is to compare the results of the two approaches. At the end an outlook and some conclusions are presented.

\section{Computational aspects}

\subsection{The numerical self-consistent-field theory}

The self-consistent-field approach is a well-known method in the field of inhomogeneous polymer systems. In principle it amounts to solving the Edwards diffusion equation in a lattice discretization. This method has already been used for obtaining the properties of a single chain in various solvent qualities \cite{edwa1965}. As a first approximation we can assume that the chain of interest is spherical symmetric, and thus the Edwards diffusion equation should be solved in this geometry. Secondly, one can consider the properties of a central chain with the constraint that segment number 1 is at the center of the coordinate system, and that the remainder of the chain is floating around this `grafting' point as a one-armed star. At the center it is then possible to account for the excluded volume of the grafting segment by not allowing other segments of the chain to enter the grafting coordinate. This leads to non-homogeneous potential fields and also non-homogeneous segment distributions of the central chain. Using this method one can obtain the swelling of the chain in good solvent and retrieve basically the Flory result that the radius of gyration is proportional to the degree of polymerization $N$ in the power 0.6 \cite{flory,secgensf}. The grafting of the end in the center is not a serious problem in the thermodynamic limit, but it is not allowed for the present class of problems. We like to know the conformational properties of relatively short copolymers and with specified primary sequence. For this reason we apply a recently developed alternative SCF procedure that yields \cite{secgensf} $R\sub{g} \propto N^{0.582}$.

Before we discuss the details we first outline the strategy of the calculations. The key idea is that not an end-point is fixed to the center of the coordinate system, but we allow any segment in the chain to be there. It is clear that not every segment in the chain has the same probability to be a central segment. Indeed, the proper statistical weight must be found for a given segment to be at the center. When some interior segment in the chain is at the center, the chain is thus modeled as a two-armed star, with two arms typically of unequal length. Thus, in our calculations we will allow only those conformations which have exactly one of its segments in the center of the system. All other possible conformations, i.e., chains that do not visit the center are disregarded. As in the one-armed star discussed above, the fact that the chain is effectively grafted in the center has the effect that the ensemble averaged chain density will be relatively high near the grafting point. This causes segments to interact with each other. In the spirit of the SCF approach, this interaction gives rise to a segment potential. This segment potential is a function of the segment densities on the one hand and it affects the segment densities on the other hand. Numerically a fixed point is found for the potential and density profiles within the constraint that the system is incompressible (all lattice sites are filled either by polymer segments or solvent molecules). Such a fixed point is the SCF solution. A typical SCF solution usually represents the equilibrium density distribution. When there is more than one SCF solution, it is necessary to analyze the thermodynamic quantities in order to select the most favorable one.

Below we give some mathematical detail of the formalism. We choose here to develop the SCF formalism by making use of lattice approximations. It is necessary to specify how this is done in order to facilitate comparison with the bond fluctuation model discussed below. We also discuss the chain propagators and briefly the analysis of the equilibrated density profiles.

\subsubsection{The lattice}

The space is discretized such that it is represented by a set of lattice sites with volume $b^3$. It is assumed that in this lattice the layers exist in which the segment density (probability that a segment of a given type occupies a site of this layer) is constant throughout this layer. In this layer we then use a mean-field approximation. In effect, the sites in such a layer are essentially indistinguishable. As the distinction of sites in the layers is unimportant it suffices to develop the equations such that only the properties of the layers occur. Differences in segment densities are allowed between layers. If the layers have spherical symmetry, there is only a gradient in the radial direction. We refer to this as a 1-gradient-SCF theory. Additionally, a 2-gradient-SCF calculation is used. The 2-gradient-SCF theory is applied in a cylindrical coordinate system: the mean field averaging is done over a ring of lattice sites. There are then gradients in density both along the long axis of the cylinder (which is the symmetry axis) as well as in the radial direction. A 2-gradient SCF theory is more powerful as it allows for cylindrical symmetric shapes of the chain such as ellipsoids, however the calculations are much more CPU intensive. In this paper, we choose to develop the theory concentrating on the 1-gradient-SCF technique. The extension to the 2-gradient-SCF technique is straightforward (see also \cite{slk1996}).

We thus consider a spherical lattice. The number of lattice sites in layers with this topology depends on the distance between the layer and the center of the system. Typically we demand the layers to have the same width of size $b$. As a consequence, we allow a layer to have a non-integer number of sites. This is not a problem because, in effect, the SCF theory is an ensemble method: the average over many copies of the central chain is computed. The first, central layer, however, is special: its radius is given by $b'$. The number of lattice sites in each layer $r = 1, ..., M$ is given by:
\begin{widetext}
\begin{equation}
    L(r) = \left\{ \begin{array}{l l} \frac{4\pi}{3}(b'/b)^3
        & \text{if}\ r=1,\\
    \frac{4\pi}{3}(((r-1)+b'/b)^3-((r-2)+b'/b)^3) & \text{otherwise}.
    \end{array} \right.
\end{equation}
\end{widetext}

In order to compute efficiently the ensemble average of all possible and allowed conformations of the chain in this lattice it is necessary to specify the {\it a priori} transition probabilities in this lattice. This probability to go from layer $i$ to $j$ is designated $\lambda(i|j)$.
\begin{equation}
    \lambda(r|r+1)=\lambda_1 A(r)/L(r)
\end{equation}
\begin{equation}
    \lambda(r|r-1)=\left\{ \begin{array}{l l} 0 & \text{if}\ r=1,\\
    \lambda_1 A(r-1)/L(r) & \text{otherwise}.
    \end{array} \right.
\end{equation}
\begin{equation}\label{eq:lambdarr}
    \lambda(r|r)=1-\lambda(r|r+1)-\lambda(r|r-1)
\end{equation}
where $A(r)$ is the outer area of a layer $r$: $4\pi(b'+(r-1)b)^2/b^2$. The parameter $\lambda_1$, which specifies the {\it a priori} probability to step from a layer to a neighboring one in the limit of $r \rightarrow \infty$ (flat layers), is a parameter which we can use to adjust the SCF theory to match the details of the MC method. The $\lambda_1$ can only be adjusted in the range were none of the transition probabilities resulting from it are less than zero.  Ensuring $\lambda(r|r)\leq0$ gives $0\leq\lambda_1\leq b'/(3b)$. In the MC system a polymer segments can have at most 26 neighbors. If the MC system would be divided in parallel layers, such an surrounded polymer segment would have 8 neighbors in the same layer and 9 in each adjacent layer. In order to have our results match the simulation data as closely as possible we have chosen $\lambda_1=9/26$ which matches the coordination number for densely packed monomers in the simulations. This means that we have $b'=27b/26$ as the smallest allowed radius for the central lattice layer.

From symmetry considerations there must exist a relation between the transition probabilities between two layers and the number of lattice sites in these two adjoining layers. It is easily shown this so-called detailed balance condition is
\begin{equation}
    \lambda(r|r+1)L(r)=\lambda(r+1|r)L(r+1)
\end{equation}
In words this equation says that there are as many ways to go from layer $r$ to $r+1$ as the other way around.

\subsubsection{The propagator formalism: from potentials to densities.}

We assume that for each segment type $x = A, B, S$, where $S$ is a solvent molecule and $A$ and $B$ are segment types of the polymer chain, there is a segment potential profile $u_x(r)$. Later on we will see that these potentials are functionals of the densities. Here we will show how the segment densities are computed from the segment potentials.

The full set of possible conformations on the spherical lattice can be split into two: all conformations that visit the central layer---we will label this set with the subscript $a$, and all other ones, which are labeled $f$. It is relatively straightforward to compute this latter set by developing the Edwards diffusion equation with the constraint that the potential felt by any of the segments is infinite in layer $r = 1$. Then no segments are allowed in this layer and the $f$-set is available. The complete set of conformations is also easily generated. The difference between the two gives the distribution of the set of conformations which we are interested in. So, our first goal is to obtain both the $f$- and the overall conformation distributions.

We introduce free segment distribution functions, defined by
\begin{equation}
    G_x(r)=\exp(-u_x(r)/kT)
\end{equation}
These segment type dependent distribution functions are generalized to free segment distribution functions that depend only on the ranking number within the polymer chain, $s = 1, ..., N$: $G(r,s) = G_A(r)q^A_s + G_B(r)(1-q^A_s)$, where $q_s^A$ is the chain architecture operator which is unity when segment $s$ is of type $A$ and zero otherwise. The set $q_s^A$ (a sequence of 0's and 1's) defines the primary sequence of the polymer chains.

The unnormalized density distribution due to all possible and allowed conformations follows from
\begin{equation}
\varphi(r,s) = G(r,s|1) G(r,s|N)/G(r,s)
\end{equation}
where $G(r,s|1)$ and $G(r,s|N)$ are chain end distribution functions which obey the Edwards diffusion equation. In the lattice approach they can be computed from the recurrence relations
\begin{equation}\label{eq:rec1}
    G(r,s|1) = G(r,s)\sidef{G(r,s-1|1)}
\end{equation}
\begin{equation}\label{eq:rec2}
    G(r,s|N) = G(r,s)\sidef{G(r,s+1|N)}
\end{equation}
where the angular brackets indicate a three layer average. Generally, $\sidef{X(r)}$ is defined as
\begin{widetext}
\begin{equation}\label{eq:sidef}
    \sidef{X(r)}=\lambda(r|r+1)X(r+1)
    +\lambda(r|r)X(r)+\lambda(r|r-1)X(r-1)
\end{equation}
\end{widetext}
where $X(r)$ is an arbitrary layer property. Here, we use $G(0,s)=0$, which means that it is impossible to penetrate to the non-existing layer $r = 0$, and $G(M+1,s)=G(M,s)$ so that between layer $M$ and $M+1$ we have a reflecting boundary condition. The recurrence relations, equations \ref{eq:rec1} and \ref{eq:rec2}, are started by $G(r,1|1)=G(r,1)$ and $G(r,N|N)=G(r,N)$, respectively.

The unnormalized density distribution of the $f$-type conformations is found as:
\begin{equation}
    \varphi_f(r,s) = \left\{ \begin{array}{l l}0 & \text{if}\ r=1,\\
     G_f(r,s|1) G_f(r,s|N)/G(r,s) & \text{otherwise}.
    \end{array} \right.
\end{equation}
with
\begin{equation}
    G_f(r,s|1) = G_f(r,s)\sidef{G_f(r,s-1|1)}
\end{equation}
\begin{equation}
    G_f(r,s|N)=G_f(r,s)\sidef{G_f(r,s+1|N)}
\end{equation}
where $G_f(r,s) = G(r,s)$ except for $r = 1$ where $G_f(1,s)=0$. Obviously the starting conditions are modified accordingly. Now, the unnormalized $a$-type conformation distribution is obtained by:
\begin{equation}
    \varphi_a(r,s)=\varphi(r,s)-\varphi_f(r,s).
\end{equation}
The normalization of the $a$-type density profile should be such that exactly $N$ segments are present in the system: $\rho(r,s)=C\varphi_a(r,s)$ with $C$ given by
\begin{equation}
    N=C \sum_{s=1}^N\sum_{r=1}^M\varphi_a(r,s)L(r).
\end{equation}
The overall density of $A$ and $B$ units is now given by: $\rho_A(r)=\sum_{s=1}^Nq^A_s\rho(r,s)$ and $\rho_B(r)=\sum_{s=1}^N(1-q^A_s)\rho(r,s)$.

The density profile of the solvent molecules is simply found as $\rho_S(r)=G_S(r)$.

\subsubsection{From densities to potentials}

From differentiation of the mean-field partition function it is possible to find an expression for the segment potentials in terms of the segment densities. This step is well documented in the literature and is not repeated here \cite{evers}. The segment potential is composed essentially of two contributions. The first one is an excluded volume potential $u'(r)$ which is a Lagrange field coupled to the requirement that each layer is exactly occupied by segments A, B, or S: $\sum_x \rho_x(r) = 1$. The second one accounts for nearest neighbor contacts parameterized by Flory-Huggins interaction parameters
\begin{equation}
    \frac{u_x(r)}{kT}= u'(r) + \sum_{x'}\chi_{xx'} (\sidef{\rho_{x'}(r)} - \rho_{x'}^b)
\end{equation}
The angular brackets denote again a three-layer average, as in eq. \ref{eq:sidef}. The segment bulk density $\rho_{x'}^b$ is 0 for segment types A and B and unity for segment type S.

\subsubsection{The SCF solution and its analysis.}

The above set of equations is closed. From the segment potentials it is possible to obtain the distribution due to the $a$-type of conformations, as well as the distribution of the solvent molecules. From the distributions it is possible to obtain the segment potentials. The value of the $u'(r)$ is adjusted such that the sum of the densities equals unity, and as soon as the segment potentials are consistent with the segment densities a fixed point is obtained. For such a SCF solution it is straightforward to analyze the total interaction energy in the system.
\begin{equation}\label{eq:Eint}
    E(T) = \frac{kT}{2} \sum_x \sum_{x'} \sum_{r=1}^M
        \chi_{xx'} \rho_x(r) \sidef{\rho_{x'}(r)}L(r)
\end{equation}
It is also easy to compute the radius of gyration
\begin{widetext}
\begin{equation}
    R\sub{g}^2 = \frac{\frac{4\pi}{5}\left(\rho(1){b'}^5
    +\sum_{r=2}^M\rho(r)((b'+b(r-1))^5-(b'+b(r-2))^5)\right)}
    {\sum_{r=1}^ML(r)\rho(r)}
\end{equation}
\end{widetext}
The extension of these formulas for calculations with two gradients is straightforward. These quantities will be presented below for a number of systems as a function of the temperature.

\subsection{The simulation technique}

In our Monte Carlo simulation we have used the well-known bond fluctuation algorithm \cite{bfm1a,bfm1b,bfm2a,bfm2b}. We consider one polymer chain of length $N$. Each monomer unit of the chain is represented as a cube taking up 8 lattice sites on a three-dimensional simple cubic lattice. The bond length $l_{bond}$ can take the values $2, \sqrt{5}, \sqrt{6}, 3, \sqrt{10}$. There are 108 different bond vectors and 87 different angles between bonds. To model the quality of the solvent an interaction potential between non-adjacent monomer units is introduced. We assign an energy $\varepsilon$ (for a homopolymer chain) to the interaction of a given monomeric unit with any other non-adjacent monomeric unit which is located in the ``first coordination cubic layer'', i.e. vectors connecting two nearest neighbors in space belong to the set $(2,0,0)$, $(2,1,0)$, $(2,1,1)$, $(2,2,0)$, $(2,2,1)$, $(2,2,2)$ including the coordinates that follow from all possible symmetry operations.

A copolymer chain consists of monomeric units of two different types A and B. Correspondingly we introduce three energy parameters for contacts between different monomeric units $\varepsilon_{AA}$, $\varepsilon_{AB}$, $\varepsilon_{BB}$, and also two parameters for interaction of monomeric units with solvent molecules $\varepsilon_{AS}$, $\varepsilon_{BS}$. It is further worthwhile to emphasize how the number of contacts between a monomeric unit and solvent are computed. We interpret each empty elementary site on the lattice as a solvent molecule. In the ``first coordination cubic layer'' there are 98 empty elementary cubes, i.e.\ 98 solvent molecules. At the same time one can easily show that due to excluded volume effects one can put maximally 26 monomeric units into the ``first coordination cubic layer'' which would correspond to the case of a pure ``polymer'' surrounding of a given monomeric unit (no solvent molecules around it). Taking this into account yields a simple linear formula for calculation of number of monomer-solvent contacts $n_{ps}$:
\begin{equation}
\label{eq:nps}
 n_{ps} = 98 \cdot\left( 1 - \frac{n_{pp}}{26} \right)
\end{equation}
where $n_{pp}$ is the number of monomer-monomer contacts.

We have studied different primary sequences of A and B monomer units with 1:1 composition (the fraction of units of each type is equal to 50\%). The so-called regular block copolymers consist of alternating blocks of A and B units of size $L= 2, 4, 5, 7, 9, 25$, respectively, and a total length of 250 monomers. The sequence of the regular block copolymers is given by (A$_L$B$_L$)$_x$A$_y$B$_y$, where $x=\lfloor125/L\rfloor$ and $y=125-Lx$. The statistical random sequences were obtained by choosing randomly the type of monomer unit at each position along the sequence.

The specially designed protein-like sequences used in the simulations and SF calculations are prepared with a coloring procedure: we collapse a homopolymer chain into a dense globule state and then examine a given snapshot. For such an instantaneous conformation, we assign the segment type A to the monomer units that belong to the surface layer and the segment type B to those that lie inside the dense core (for details of this procedure see our previous papers \cite{paper1,paper3,paper5}). The protein-like sequences were shown to have long range correlations along the chain \cite{paper7}. It was also shown that the primary sequence has `memorized' some of the features of its `parent' conformation \cite{paper5}.

The standard motion of the chain is implemented by local jumps of monomer units by one lattice site in the 6 possible lattice directions. One Monte Carlo step (MCS) corresponds to $N\,$ attempted monomer unit moves. The moves are accepted according to the usual Metropolis rules.

The simulations are performed in the following way. For each temperature, we start from a self-avoiding walk configuration and let the chain equilibrate during 10 million MCS.  Interaction parameters were fixed and will be discussed below. The averaging was performed over $M$ runs with independent starting self-avoiding walk configurations. Typically $M$ was equal to 20. The number of MC steps of the simulation runs is chosen in such a way that during the second half of the run (the next 10 million MCS) the mean values of measured quantities did not differ from the ones found in the first half of the run within the error of measurement. Such procedure, however, cannot exclude freezing in metastable conformations at low temperatures, where the acception rate of elementary moves goes down drastically. This is the reason why we have performed computer simulations only in the vicinity of the coil-globule transition temperature and not at temperatures far below this transition temperature.

For a given sequence we run this procedure for a number of temperatures in order to find out where the transition region is. Then we collect a detailed energy and gyration radius histogram at these points in the transition region and perform multiple histogram reweighting \cite{histrw,ipb1998} to obtain a reliable estimate of the full transition curve.

\subsubsection{Parameter choice and the temperature scale}

The central idea behind our investigation is to obtain information about the globule to coil transition in copolymers with pre-set primary sequences. In a number of steps with increasing complexity in the system we hope to obtain information about the stability of protein-like molecules. The first step in such a process is to consider copolymers composed of just two type of segments. The parameter choice is chosen such that there are polar and apolar segments.

In the MC simulations the following parameter set was used: $N = 250$, $\varepsilon_{SS} = 0$, $\varepsilon_{AA}= 0$, $\varepsilon_{BB} = -1$ kT, $\varepsilon_{AS} = -26/98$ kT, and $\varepsilon_{BS} = 26/98$ kT. Here we assume that these parameters are purely enthalpic in nature. The reason for the special quantities for the last two parameters will become clear when we map this parameter choice onto FH $\chi$-parameters that are needed in the SCF calculations. Without losing generality we set the Boltzmann constant $k = 1$, which fixes the temperature scale, and use the numerical values as mentioned above for $T = 1$. To obtain the relevant FH parameters we consider first a system of monomer units of two types in both models. The complication that arises is that the co-ordination number is not a constant in the MC simulations, i.e., it depends on the type of segments surrounding another segment. Keeping this in mind we write
\begin{equation}\label{eq:chi}
    \chi_{xy} = \frac{1}{kT} \left[ Z_{xy} \varepsilon_{xy}
-\frac{(Z_{xx}\varepsilon_{xx}+Z_{yy}\varepsilon_{yy})}{2} \right]
\end{equation}
where $Z_{xy}$ is the coordination number of $x$ units surrounding $y$ units. From the above we know that there are two possible coordination numbers in the system: it takes the value 26 for polymer segments surrounding other polymer segments densely, and 98 when only solvent molecules are involved in the nearest neighbor contacts around a polymer segment. Thus, $\chi_{AB} = [26\cdot0-(26\cdot(-1)+26\cdot0)/2]= 13$, $\chi_{AS} = [98(-26/98)-(26\cdot0+98\cdot0)/2] = -26$ and $\chi_{BS} = [98(26/98)-(26\cdot(-1)+98\cdot0)/2] = 39$. Again, the specified FH interaction parameters will be used for $T = 1$, and they will be modified when the temperature differs from this value. To mimic the simulation model as closely as possible in the SCF calculations we set $\lambda_1=9/26$ and $b'/b = 27/26$.

One may argue that in the SCF model, where the first-order Markov approximation was used in the chain statistics, a lower number of beads along the chain should be chosen in order to correctly compare to the MC simulations. However, as we like to perform the calculations on exactly the same primary sequences as was done in the MC simulations, especially as it comes to the protein-like primary sequences, we have chosen to use $N = 250$ also in the SCF calculations.

\section{Results and discussion}

Results of both SCF calculations and MC simulations will be presented for exactly the same polymer sequences. First of all there are the regular block copolymers with the generic structure (A$_L$B$_L$)$_x$A$_y$B$_y$, where $x=\mathrm{int}(N/2L)$ and $y=N/2 - xL$. Indeed, the so-called protein-like sequences as generated by the MC technique mentioned above were also used in the SCF calculations. In principle we are interested in the generic differences between the designed copolymers and the random ones. To eliminate anomalies due to uniqueness of particular sequences, we have also averaged over 17 random and over 17 protein-like sequences, respectively. We will start with presenting the SCF results, then discuss the MC data and compare those to the SCF ones.

\subsection{SCF results}

The main point of interest is the behavior of the copolymer chains as a function of the temperature. In particular we will discuss the interaction energy per segment and the size of the molecules. In the SCF results the average interaction energy per segment is normalized to the value at high temperature and thus this quantity approaches zero at high temperature. In this way we are able to return to a similar energy scale as in the MC simulations.

\begin{figure}
\resizebox*{0.75\columnwidth}{!}{\includegraphics{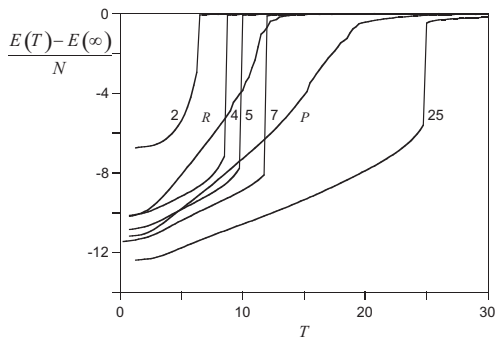}}
\caption{\label{fig:SFinten} The temperature dependence of interaction energy
per monomer for the regular (labeled by a number), protein-like ($P$) and random ($R$) copolymers as obtained from SCF calculations and given by eq.~\ref{eq:Eint}. The Flory-Huggins interaction parameters are assumed to be enthalpic quantities.  The regular block-copolymers are labeled with their value for $L$. The $P$ and $R$ curves are the average of the results for 17 different protein-like and random sequences, respectively.}
\end{figure}

\begin{figure}
\resizebox*{0.75\columnwidth}{!}{\includegraphics{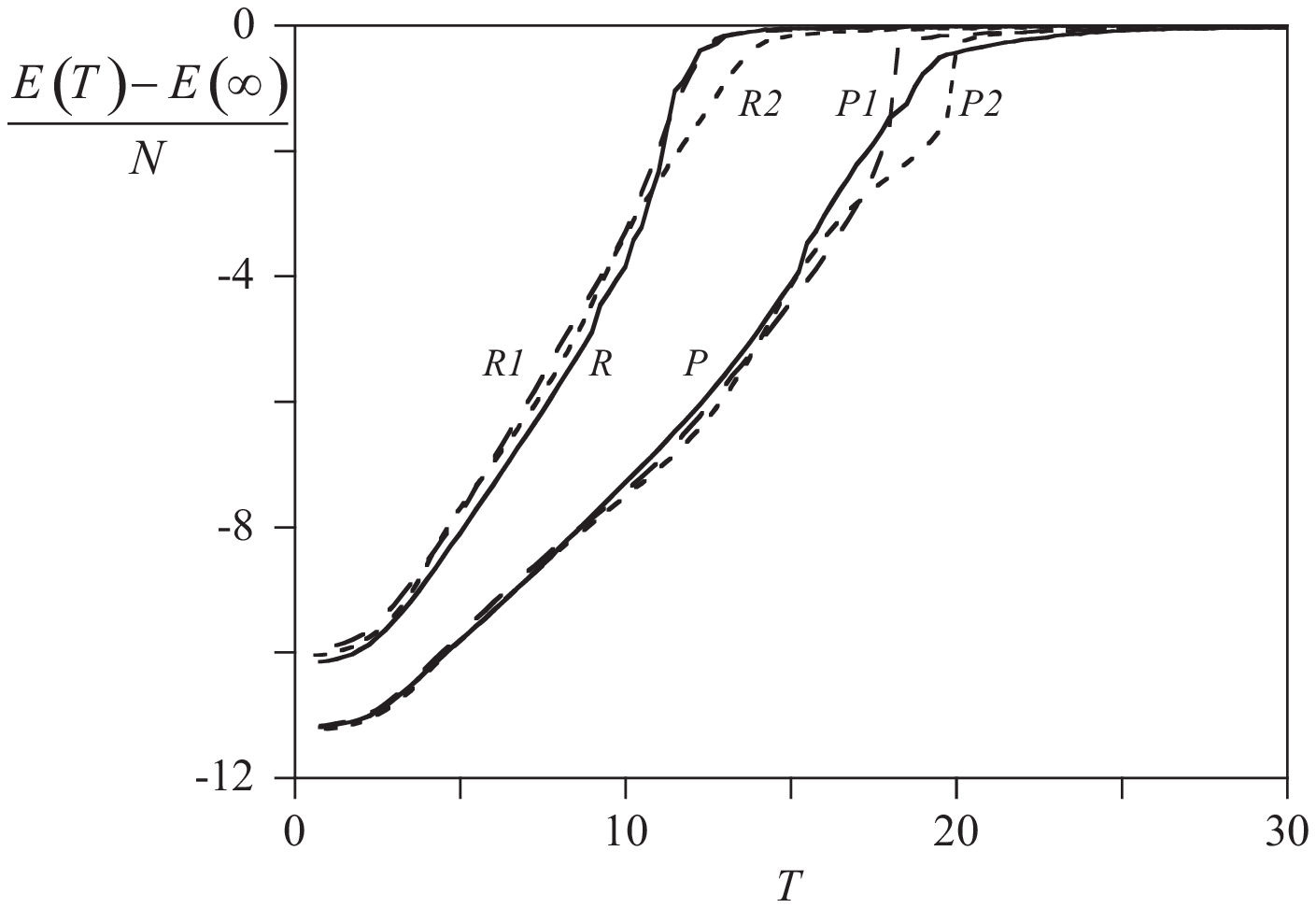}}
\caption{\label{fig:sfavgeint} This plot is similar to figure \ref{fig:SFinten} but only curves for random and protein-like sequences are plotted. Individual random sequences are indicated by $R1$ and $R2$. Similarly, $P1$ and $P2$ denote curves for protein-like sequences. The curves $P$ and $R$ (averaged over 17 sequences) are taken from figure~\ref{fig:SFinten}.}
\end{figure}

In figure \ref{fig:SFinten} the normalized interaction energies per segment are plotted as a function of the temperature $T$ for the primary sequences used. Figure \ref{fig:sfavgeint} shows more detail for the random and protein-like sequences. Here we have assumed that the interaction parameters in the SCF model are inversely proportional to $T$ and thus that they are purely enthalpic. The random and protein-like curve are averaged over 17 different primary sequences. In figure \ref{fig:sfavgeint} the two curves are re-plotted together with two typical examples of particular results found for two specified sequences. It is seen that the individual curves deviate only a little bit from the average. In all cases investigated the deviations from the average were significantly smaller than the difference between the random and the protein-like sequences. Inspection of figure~\ref{fig:SFinten} and figure~\ref{fig:sfavgeint} shows that it is possible to distinguish four regions in each curve.

\begin{description}

\item[(i)] At low temperature, in the collapsed state, the energy is constant and this state may be called the ground state. With increasing block size $L$ for the regular copolymers we notice that the ground state energy decreases. The reason for this is that small blocks along the chain can not avoid internal A-B contacts. The internal structure of the $L = 2$ globules is a layered one. The molecule core consists of alternating (lattice) layers filled by A and then by B units. For larger block lengths the regions rich in A and B are larger than a single lattice layer and therefore the ground state energy can become lower. The ground state level of a random sequence is significantly higher than that for a protein-like chain. This is because in the protein-like sequence there are longer stretches of apolar units which can efficiently pack in the interior of the dense globule than in the random sequences. Here we thus notice that the coloring procedure results in chains that significantly differ from their random analogues.

\item[(ii)] At slightly higher temperature, fluctuations in chain conformations
become possible and the compact molecule undergoes internal rearrangements. As we will see below, the chains remain rather compact but the interaction energy per segment increases more or less linearly with the temperature $T$. We refer to this regime as the molten globule state \cite{moltglob}. There is not a strong $L$ dependence for the onset of the molten globule domain. When more entropic restrictions are locked in into the ground state (i.e.\ for smaller $L$), the system enters the molten globule state easier. The longer the block length the larger the temperature range over which the molten globules state is found. Both the protein-like and the random copolymers have a well-developed molten globule regime. For protein-like chains this region is wider than the corresponding region for a random copolymer chain, again because the latter has, on average, shorter stretches of apolar chain segments.

\item[(iii)] At the end of the molten globule state there is a sudden jump in the
interaction energy. At this temperature $T\sub{trans}$ the coil to globule transition is found. In several cases, especially for the regular polymers with large blocks, the transition appears as a first-order-like transition. The transition temperature increases with increasing block length. For the regular block copolymers the jump appears especially large for chains with intermediate block lengths. Below we will comment on the cooperativity of the transition in relation to the assumptions made in the SCF theory. Significantly, we find that the protein-like and the random copolymers have a very small jump (if any) at the coil to globule transition. The transition thus tends to become continuous with increasing levels of heterogeneity along the chain.

\item[(iv)] Finally, at very high temperature the chains are swollen and behave as
a random (or even swollen) coil. In such a chain the majority of the interactions are between polymer units and solvent molecules. This state does not depend on the primary structure of the chains and therefore the interaction energy levels off to a constant value, which due to the normalization is equal to zero.

\end{description}

\begin{figure}
\resizebox*{0.75\columnwidth}{!}{\includegraphics{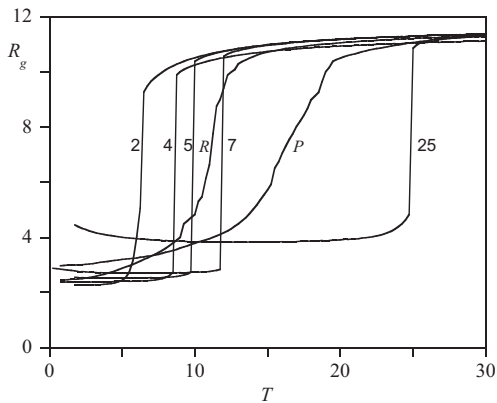}}
\caption{\label{fig:SFRg} The radius of gyration versus the temperature $T$ as found in the SCF calculations also plotted in figure~\ref{fig:SFinten}. The labeling of the curves is similar as in figure~\ref{fig:SFinten}. The Flory-Huggins interaction parameters were taken to be enthalpic quantities.}
\end{figure}

\begin{figure}
\resizebox*{0.75\columnwidth}{!}{\includegraphics{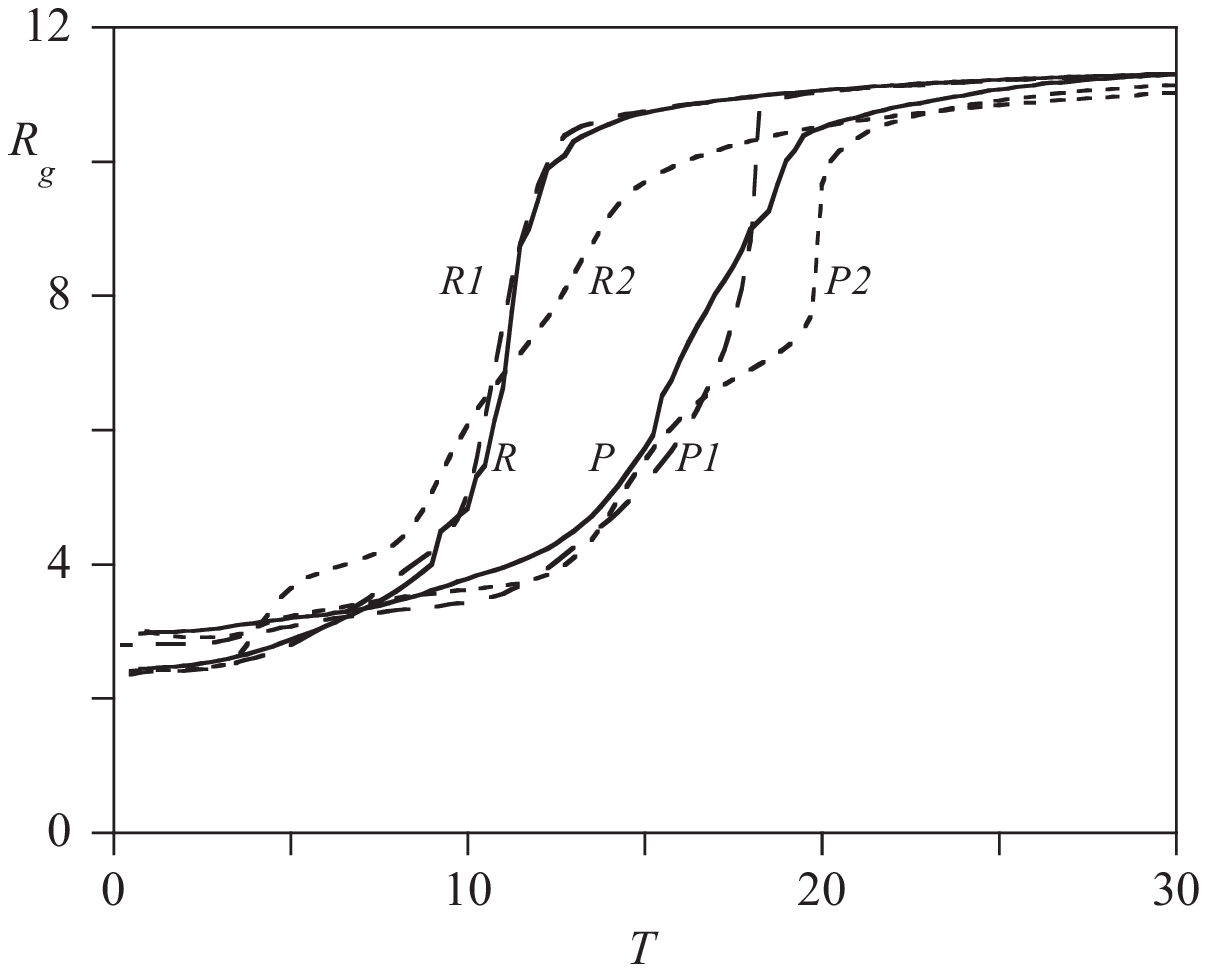}}
\caption{\label{fig:sfavgrg} As figure~\ref{fig:SFRg}, but now only for the random and protein-like sequences, both the averages and the individual sequences are from the same computations as in figure \ref{fig:sfavgeint}.}
\end{figure}

In figure \ref{fig:SFRg} and figure \ref{fig:sfavgrg} we collect the corresponding radii of gyration as a function of the temperature $T$. Again the corresponding curves plotted in figure \ref{fig:sfavgrg} show that the deviations that exist within one sequence type is small as compared to the deviations in the shape and position of the curve between sequence types. We have corrected both the SCF results and the computer simulation results (see below figures \ref{fig:HRRg} and \ref{fig:HRRg2}) for the radius of gyration to a system with $Z=6$ and bond length equal to segment diameter. The relation $R\sub{g} \propto\sqrt{\lambda_1}$\cite{janenskortsov} gives us that the $R\sub{g}$'s found with the SCF method should be multiplied by $\sqrt{(\frac{26}{9}\frac{1}{6})}$. The $R\sub{g}$'s from the MC simulations have to be divided by the average bond length: $2.7$. Typically one would expect the chains to have the largest size at high temperature, where the chains are swollen coils, and the smallest size at low temperature (ground state). Inspection of figure \ref{fig:SFRg} shows that, as expected, all curves tend to the same coil size for large temperature. There is a gradual change in $R\sub{g}(T)$ in the swollen regimes. For small $L$ the curves are rounded whereas for larger $L$ the limiting value is found almost directly after the coil to globule transition. At the coil to globule transition temperature the size decreases sharply upon decrease of $T$. For the regular block copolymers the coils collapse almost immediately to a compact size. This fact can be understood as a consequence of effective stiffness of a copolymer chain: one can represent a regular multiblock copolymer chain as a rather stiff homopolymer chain with a large and highly anisotropic effective monomer unit equivalent to one full $A_L B_L$ block in the original copolymer chain. This effective stiffness due to renormalization of the monomer units leads to more sharp coil-globule transition \cite{book}. The distinction between the ground state and the molten globule is not clearly visible from inspection of the radius of gyration. This is in agreement with the notion of a molten globule as a dense weakly ordered structure \cite{moltglob}. The size at low temperature decreases with decreasing block length. The reason is that a structure formed by a regular block copolymer with large block length $L$ tends to be a collapsed core with a swollen corona with solvated loops. For small $L$ no loops can develop. Interestingly, for $L = 25$ a minimum in the radius of gyration is found as a function of the temperature. This minimum is attributed to the miscibility of A in B (and B in A) at intermediate temperatures. Due to this the length of the loops tend to go through a minimum. For the random and the protein-like sequences, the radius of gyration increases relatively smoothly with the temperature. The variation in block lengths present in both the random and the protein-like sequences is responsible for the more gradual increase of the size of the molecule.

The cooperativity of the coil to globule transition as seen in the SCF calculations is somewhat overestimated. In figure \ref{fig:SFinten} it was noticed that for $L= 5$ the jump in energy is particularly large. Also the size increase at the transition (cf.~figure \ref{fig:SFRg}) is very pronounced for this case. Due to the mean field approximation in spherical geometry, it is necessary that during the transition the chains remain spherically symmetric. Therefore the transition tends to be very cooperative; the entire chain has to go collectively through huge conformational changes. In more realistic cases the chain can locally swell or collapse and therefore the cooperativity of the transition tends to be lower than predicted by the classical SCF approach. In a 2-gradient SCF approach as mentioned in the theoretical section, it is possible that the coil to globule transition occurs locally. The 2-gradient SCF results were obtained in a cylindrical coordinate system. As in the 1-gradient SCF theory the chain is forced to go through a pre-assigned coordinate which we denote as the origin $(0,0)$ at the central lattice layer on the symmetry axis of the lattice. In addition the position of one of the ends of the chain is restricted to coordinates $(z',0)$ where $z'>0$ is a coordinate on the symmetry axis of the coordinate system. In this way rotational degeneracy of chain conformations is partly removed from the calculations.

\begin{figure}
\resizebox*{0.75\columnwidth}{!}{\includegraphics{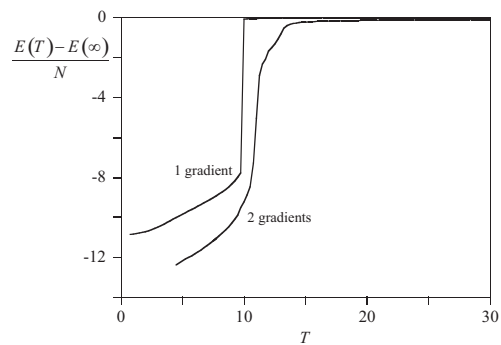}}
\caption{\label{fig:1and2G} Normalized segmental energy as a function of the
temperature $T$ for a regular block copolymer $L=5$ as found by SCF calculations for the 1-gradient and 2-gradient approaches as indicated.}
\end{figure}

The impact of the extra degree of freedom in a 2-gradient SCF approach is illustrated in figure \ref{fig:1and2G} for the most pronounced case $L = 5$. In figure \ref{fig:1and2G} the normalized energy per segment as a function of the temperature $T$ is plotted for the two versions of the SCF theory. As anticipated the transition is continuous in the 2-gradient case. Note as well that the transition temperature is not the same in both types of calculations. The reason is that the extra degree of freedom allows a more efficient lowering of the free energy of the coils. In other words the molecules tend to be more stable at any temperature. This is thus also reflected in the increase of the transition temperature. The conclusion is that the coil to globule transition is likely to be a continuous transition for this type of regular copolymers.

\subsection{Computer Simulation Results}

We discuss the results of Monte Carlo computer simulation for copolymers with different primary sequences. Because it is impossible to sample reasonable statistics in dense globular state using local moves only, we have studied the transition region and performed the histogram reweighting to slightly lower temperatures. In our simulation we can calculate, e.g., the energy, the gyration radius, the mean cluster size (number of monomeric units of type B which belong to the dense core) and the fluctuations of all these quantities. For the sake of comparison we concentrate here on the first two of these quantities.

\begin{figure}
\resizebox*{0.75\columnwidth}{!}{\includegraphics{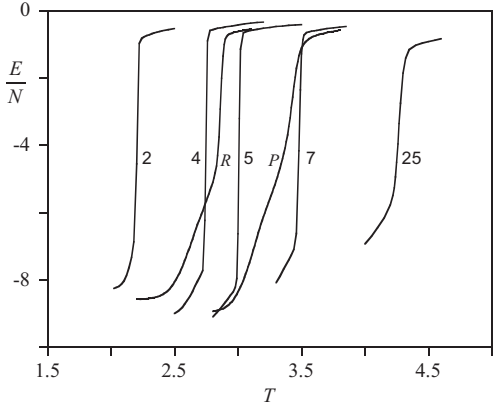}}
\caption{\label{fig:HRinten} The interaction energy per monomer unit
for all primary sequences is plotted versus the system temperature (MC computer simulation results). Regular block copolymers are labeled with the value for $L$. The protein-like $P$ and random $R$ curves were obtained by averaging the data obtained for 17 different sequences.}
\end{figure}

\begin{figure}
\resizebox*{0.75\columnwidth}{!}{\includegraphics{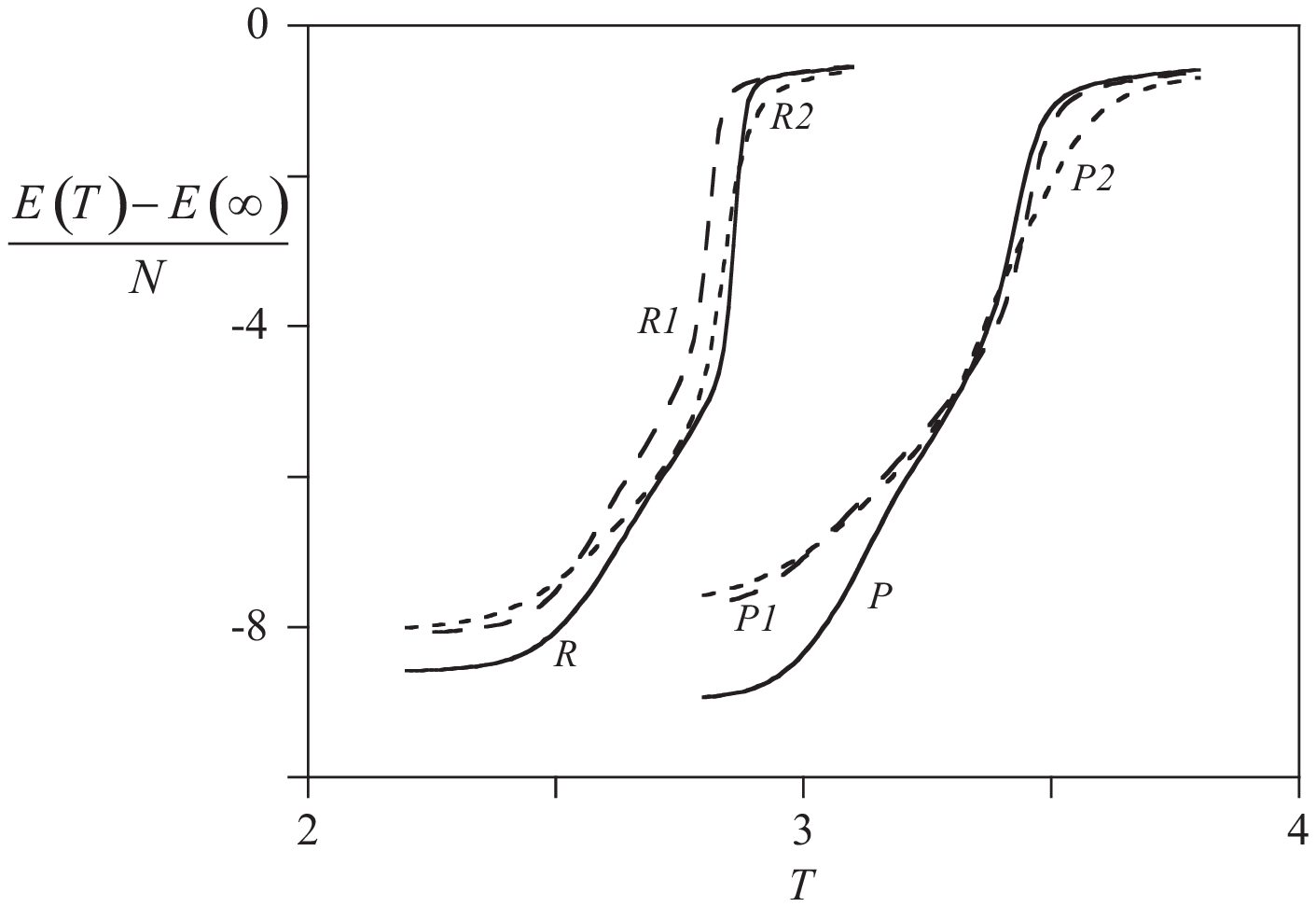}}
\caption{\label{fig:HRinten2} This plot is similar to figure~\ref{fig:HRinten} but only curves for random and protein-like sequences are plotted. Individual random sequences are indicated by $R1$ and $R2$.
Similarly, $P1$ and $P2$ denote curves for two particular protein-like sequences. The curves $P$ and $R$ (averaged of 17 sequences) are taken from figure \ref{fig:HRinten}.}
\end{figure}

In figure \ref{fig:HRinten} the segmental energy is plotted as a function of the temperature $T$ for the regular block copolymers, the averaged random and averaged protein-like sequences, while in figure \ref{fig:HRinten2} the results of the simulations are shown for the individual random and protein-like primary sequences identical to those in figure \ref{fig:sfavgeint}. Completely in line with the SCF results we find that the transition temperature increases with block size $L$. All curves in this figure tend to go towards $E/N = 0$ for large temperatures. This is the natural consequence of the choice of the parameters: the A and B monomers have interaction potentials with the solvent that are equal in magnitude but differ in sign. So unlike for the SCF calculations, a normalization to high temperatures is not needed.

The histogram reweighting results are expected to be accurate only in the vicinity of the transition region, but not at very low temperatures $T$. The reason for the latter is that a rare event of finding a low energy while simulating in the transition region determines the energy level of the ground state. This then results in an inaccurate estimate for this limit, because the true ground state level was probably not visited at all during the simulation. The molten globule regime as found in SCF calculations is relatively more difficult to retrieve accurately in Monte Carlo simulation because of the relatively poor statistics for compact globules. Moreover, the conformations of compact globules obtained in MC simulation at low temperatures correspond to the case of the so-called crumpled globule and not to the equilibrium one \cite{crumpl}.

With this in mind, we notice that the shapes of the $E(T)$ curves in figure \ref{fig:HRinten} are qualitatively similar to those found by SCF calculations (figure~\ref{fig:SFinten}). In line with the SCF results the block copolymers have a more cooperative coil to globule transition than the random and protein-like sequences. Also, the transitions of the protein-like sequences are more gradual than those of the random ones. Finally, good agreement between SCF and MC results is found for the order of the transition temperature for the random and protein-like sequences in relation to the regular block copolymers. Interestingly, also the difference in interaction energy per segment between the high and low temperature limits is not quite the same in both methods. Apparently, the number of contacts between different type of segments in the ground state is not the same in both methods.

One of the differences between both methods is the much shorter molten globule regime in the MC results. In the SCF results, the interaction energy was found to increase linearly with the temperature in this regime. In MC results this regime can be indicated for regular copolymers and is better visible for random and protein-like copolymers, but it is sharper and shorter than that in SCF. We can not decide at the moment what the reason is for this disagreement: whether it is only due to the relatively poor statistics of the MC method in this regime, or also due to an inherent problem in the SCF calculations. Another problem, obviously, is the huge difference in the temperature scale between both computational methods. We return to this point below.

\begin{figure}
\resizebox*{0.75\columnwidth}{!}{\includegraphics{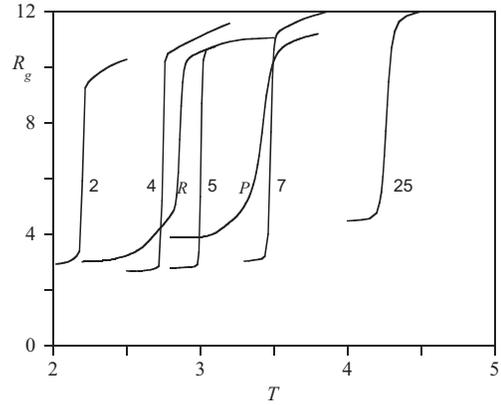}}
\caption{\label{fig:HRRg} The radius of gyration versus the temperature $T$ as found in the MC simulations also plotted in figure~\ref{fig:HRinten}. The labeling of the curves is similar as in figure~\ref{fig:HRinten}.}
\end{figure}

\begin{figure}
\resizebox*{0.75\columnwidth}{!}{\includegraphics{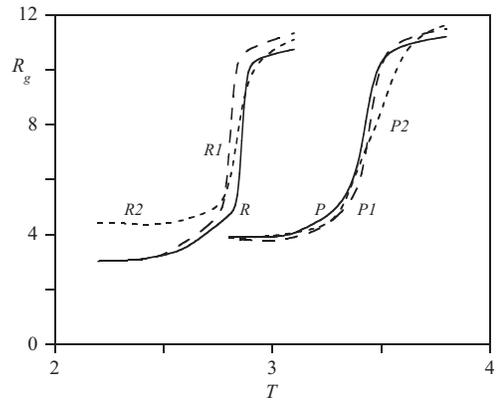} }
\caption{\label{fig:HRRg2}As figure~\ref{fig:HRRg}, but now only for the random and protein-like sequences, both the average and the individual sequences are from the same computations as in figure~\ref{fig:HRinten2}.}
\end{figure}

As to the radii of gyration, the counterpart of figure~\ref{fig:SFRg} and figure~\ref{fig:sfavgrg} are figure~\ref{fig:HRRg} and figure~\ref{fig:HRRg2}, respectively. Quantitatively, the results of both methods are again in good agreement for both the collapsed and coil state. With respect to the coil-globule transition, it is of interest to point to a few pertinent features in the MC plots which compare favorably with the SCF results. One of these is the value of the radius of gyration at the temperature just above $T\sub{trans}$. This value decreases significantly with block size $L$; more generally it decreases with decreasing transition temperature. This may be attributed to the fact that for chains that have many A-B connections along the chain (e.g., small $L$), the (renormalized) solvent quality just above the transition is still rather poor. It takes a significant temperature increase before the good solvent regime is reached. Clearly these aspects are present in both computational methods. Not surprisingly the size of the globules is only a weak function of the primary sequence. As a consequence the jump in radius of gyration decreases with block size $L$. Again, in line with the findings in the SCF calculations, the presence of long A loops, possible in the primary sequence with high $L$, results in a slightly higher $R\sub{g}$ of the ground state.

The protein-like and random sequences show also in the $R\sub{g}(T)$ plot a less steep increase at the transition temperature. This is completely in line with the decreased cooperativity of the transition as found from the interaction energy curves discussed above. Upon close inspection however, the slope in the transition region for the protein-like copolymer is approximately half of that for the random sequence. This is in agreement with SCF calculations, although the SCF results show this trend to be more pronounced. Again this may be attributed to the differences in the temperature scale as we will show below. The deviation within a type of primary sequence as plotted in figures~\ref{fig:HRinten2} and \ref{fig:HRRg2} proves that the simulations and the SCF calculations are both sensitive in a comparable way to the primary sequence. This shows that both computation tools are consistent and have predictive power.

\begin{figure}
\resizebox*{0.75\columnwidth}{!}{\includegraphics{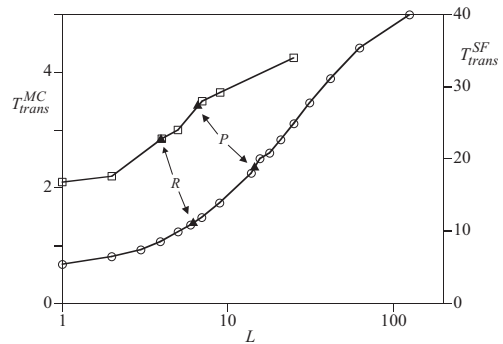}}
\caption{\label{fig:ttrans} The transition temperature $T\sub{trans}$ as a
function of the length of the blocks $L$ as found in MC simulations (open square, left ordinate) and in  SCF calculations (open circles, right ordinate). In the SCF results the temperature is defined by taking the Flory-Huggins interaction parameters to be enthalpic quantities. The transition temperature for the random sequences and the protein-like sequences as found with the respective methods are indicated by the filled triangles.}
\end{figure}

It is of interest to investigate the transition temperature as a function of the block lengths as found in both methods. We have determined the transition temperature as the temperature where the heat capacity
%(energy fluctuation)
\begin{equation} \label{fe1} C_V = \frac{<E^2> - <E>^2}{k_B T^2}
\end{equation}
has its maximum. From the curves in figure~\ref{fig:SFinten} transition temperatures can be found by taking the maximum derivative of the curves. In figure~\ref{fig:ttrans} these quantities are collected. In this graph we included block lengths up to $L = 125$ in the SCF calculations. For the SCF computations a sigmoidal shape is found on a linear-log plot. Although much less clear, this shape of the curve is consistent with the MC data. At small $L$ value the transition temperature is relatively independent on $L$. For intermediate $L$ value we find that the transition temperature scales logarithmically with $L$. This regime is lost for very high $L$ where there are just two blocks in the molecule.

In figure \ref{fig:ttrans} the transition temperatures of the random and protein-like sequences are also included. The random sequences show a transition temperature which is equivalent to the $L = 4$ regular chains as found by means of MC simulation and to slightly higher value $L=6$ in SCF approach. We note that the number average block size of random sequences is just 2. Apparently, the transition temperature is determined by the longer block lengths. A similar phenomenon is observed for the protein-like sequences. Again, the transition temperature is much higher than could be expected from the number average block length: the number average length is around 3.7 whereas the transition temperature occurs around the $L = 7$ (MC results) or $L=12$ (SCF results) position for regular chains.

Quantitative agreement for the temperature scales can not seen because there remain significant differences between the two computation techniques. Ideally we would have a temperature in both models that can be directly compared. However, there are a few intricacies which can be traced back to fundamental differences between the two methods.

\begin{description}

\item[(i)] In the MC simulations the through-bond contacts are irrelevant for the statistical weight of a given conformation. For this reason these bonds are not counted. In the SCF model as specified above, the energetic contacts are counted in the system as if the segments are detached from each other. The average surrounding of each segment is counted as if the bonds were not there. Of course this is not correct, but it simplifies the calculations. Typically this error is considered to be a minor point in SCF models and therefore this approximation is usually accepted. In principle one can rigorously correct for this in a mean field model. Alternatively, one can effectively deal with it by renormalizing the co-ordination number used in the calculations. Equivalent to this it can be dealt with by redefining the FH parameters that are in the system at $T = 1$. Recall that the number of possible segment-segment contacts in the simulations is 26, and at most two of these are through-bond contacts. For this reason we did not incorporate a correction and accept a small change in the temperature scale. 

\item[(ii)] The second problem originates from the fact that in the bond fluctuation model it is possible that the bond length varies, whereas in the SCF model this is fixed. As in the bond-fluctuation model there is no energetic penalty for the variation in bond length, it will tend to be large in good solvent conditions, and relatively small in bad solvent conditions. The bond length variations are a seat of entropy not present in the SCF model. To correct for this, we should write the FH parameter as a free energy parameter.

\item[(iii)] The third problem is due to the difference in size between a polymer segment and a solvent molecule. One may argue that in the above translation between the energy parameters as used in MC to the FH parameters this was accounted for. However, there remains the problem that there must be some entropy associated to the difference in size between these units. Again, this problem can in first approximation be lifted by suggesting that the FH parameters are free energy parameters.

\end{description}

Especially the latter two problems seriously frustrate our attempts to compare both methods quantitatively as to the temperature where the transitions take place. In an attempt to do this more quantitatively one has to estimate how the three FH interaction parameters are to be split up into enthalpic and entropic parts. Both problems under consideration are not likely to affect the AB interaction parameter. This one can be considered enthalpic as it is. The other two however must be corrected by a $-\Delta S$ contribution. One should realize however that points (ii) and (iii) are related: when the solvent molecules are of the same size as the polymer units, it would be hard to come up with a method that allows the bonds to fluctuate without introducing serious packing problems. In a densely packed phase with polymer units (the reference) there are just 6 bond directions (all of length 2). Alternatively in the dilute case there are 108 bonds and therefore the correction should be of the order of -ln(108/6). With this correction we can compute the FH parameters as a function of the temperature:
\begin{subequations}
\label{eq:TfreeEn}
\begin{equation}
\chi_{AB} = 13/T
\end{equation}
\begin{equation}
\chi_{AS} = -26/T-\ln(108/6)
\end{equation}
\begin{equation}
\chi_{BS} = 39/T-\ln(108/6)
\end{equation}
\end{subequations}
Because of the intrinsic differences we have to accept some disparity in temperature scales in between both methods. We expect that a better comparison of the temperature scale can be obtained by using eq.~\ref{eq:TfreeEn} to calculate the interaction parameters.

\section{Outlook}
From the above it is clear that the two computational techniques give qualitatively similar results for the coil to globule transition of AB copolymers. Both techniques have strong points, but also have their weaknesses. Although in principle the MC technique is more exact, in practice computational restrictions often impede the theoretical advantage. This is particularly evident when information is needed for compact globules. The SCF method is computationally very inexpensive but has the intrinsic problem that it is impossible to investigate individual conformations. Clearly the methods complement each other. This means that one can use SCF calculations to investigate the sequence and parameter space efficiently and save computer time such that MC simulations can be performed on more interesting systems.

As there is qualitative agreement on the behavior of the coil to globule transition in the bulk, it will be of significant interest to put both methods to the more challenging problem of AB copolymers at interfaces. This problem has several aspects. First the adsorption of the chains from the bulk onto the surface needs to be investigated. For this it will be important to measure the free energy as a function of the distance from an adsorbing interface. Secondly, an interface often induces structural changes or unfolding transitions. These and other aspects are now subject of investigation.

\section{Conclusions}

We have analyzed the coil-globule transition of copolymer systems by two computationally very different methods namely SCF calculations and MC simulations. The two methods complement each other. The SCF technique is computationally fast and captures the qualitative features rather well, as proven by the more exact MC results. Due to the differences in the methods there exists a fundamental problem with comparison of the temperature scales in both methods. We have not studied here the finite size effects \cite{fss} so that the extrapolation to infinite chain length is not yet available.

Detailed information is obtained for the coil to globule transition of copolymers with two type of segments. The transition temperature was found to be roughly an exponential function of the block length of regular copolymers. The type and number of segment-segment contacts in the ground state were recovered in SCF method showing that this quantity is well defined for a particular copolymer primary sequence. It was not possible to sample the ground state in MC simulation using only local moves. The cooperativity of the transition has been shown to be high for regular block copolymer molecules. This is explained by renormalization of effective monomer unit \cite{book} in the case of monodisperse blocks. The cooperativity is lower for the random and protein-like sequences because of the polydispersity of the block length. Further, it was shown, in agreement with previous results \cite{paper1,paper5}, that the globular state for protein-like sequences is more stable than that for statistical random copolymers. Beside a ground state and an unfolded coil state we have recognized a molten globule state which appears most pronounced for the protein-like sequences. Since protein-like sequences behave randomly, the calculations thus confirm the previous findings \cite{paper1,paper5} that the protein-like molecule can inherit some information from the parent conformation used to define its sequence.

\section*{Acknowledgments}

The financial support from NWO (grant No. 326-758), from INTAS (grant No. INTAS-97-0678) and from Russian Foundation for Fundamental Research is kindly appreciated. The authors thank K.B.Zeldovich both for stimulating discussions and extensive technical support.

\end{document}